\documentclass{aa}
\usepackage{graphicx}
\def\lax    {\ifmmode{_<\atop^{\sim}}\else{${_<\atop^{\sim}}$}\fi}
\def\gax    {\ifmmode{_>\atop^{\sim}}\else{${_>\atop^{\sim}}$}\fi}
\def\kms    {\ifmmode{{\rm ~km~s}^{-1}}\else{~km~s$^{-1}$}\fi}

\def\approx   {$\sim$}
\def\arcm   {$^{\prime}$}
\def\arcmper  {\ifmmode \rlap.{' }\else $\rlap{.}' $\fi}

\def\arcs   {$^{\prime\prime}$}
\def\arcsper  {\ifmmode \rlap.{'' }\else $\rlap{.}'' $\fi}
\def\arcsgper  {\ifmmode \rlap.^{s }\else $\rlap{.}^s $\fi}

\def\deg      {\ifmmode^\circ\else$^\circ$\fi}     

\def\hper     {\ifmmode \rlap.^{h}\else $\rlap{.}^h$\fi}

\def\m1       {$^{-1}$}

\def\mper     {\ifmmode \buildrel m\over . \else $\buildrel m\over.$\fi}

\def\sper     {\ifmmode \rlap.^{s}\else $\rlap{.}^s$\fi}

\def\>           {$>$}
\def\<           {$<$}

\def\simlt       {\lower.5ex\hbox{$\; \buildrel < \over \sim \;$}}
\def\simgt       {\lower.5ex\hbox{$\; \buildrel > \over \sim \;$}}

\begin{document}
%
%
\title{Star formation in the warped outer pseudoring of the spiral 
galaxy NGC 3642}
\author{
 L. Verdes--Montenegro\inst{1} \and A. Bosma\inst{2} \and
 E. Athanassoula\inst{2}}
\institute{Instituto de Astrof\'{\i}sica de Andaluc\'{\i}a, CSIC,
Apdo. 3004, 18080
Granada, Spain \and 
Observatoire de Marseille, 2 Place le Verrier, 13248 Marseille 
Cedex 4, France}
\offprints{L. Verdes--Montenegro}
\date{Received ; accepted }

\titlerunning{The warped galaxy NGC 3642}
\authorrunning{Verdes-Montenegro et al.}
\abstract{NGC 3642 was classified as a spiral
galaxy with three rings and no bar.
We have performed an HI and optical study of this nearly face-on
galaxy. We find that the nuclear ring might in fact be
part of an inner one-armed spiral, that could be driving nuclear
accretion and feeding the central activity in the inner kpc.  The
inner ring is faint, and the outer ring is a rather ill-defined
pseudoring.  Furthermore, the size ratio of the rings is such
that they cannot be due to a single pattern speed linking them
together.

The outer pseudoring is peculiar, since it lies 
in the faint  outer parts of
the disk, where star formation is still going on
at 1.4 times the optical radius.  Higher HI column densities are
associated with these regions and the atomic gas layer is warped. 
These perturbations affect only the
outer disk, since the kinematics within the main body conforms well
to
an ordinary differentially rotating disk. 

We propose here that both 
nuclear activity and star formation in the warped outer parts
might be linked to the fact that  NGC 3642 is located in a rich
environment, 
where its close neighbors show clear signs of merging.
Our suggestion is that 
 NGC 3642 has
captured recently a low-mass, gas-rich dwarf, and star formation 
was triggered in this infalling 
external gas that produced also a pronounced warp
in the gaseous disk.}
\keywords{galaxies: individual: NGC 3642 - galaxies: kinematics - galaxies: 
photometry - galaxies: spiral - galaxies: structure}
\maketitle

\section{Introduction}

NGC 3642 was classified as unbarred (SAbc from RC3 - De Vaucouleurs et
al. 1991, Sb(r)I from the RSA - Sandage \& Tammann 1981) and has three
rings (De Vaucouleurs \& Buta 1980, hereafter DVB80).  This
combination was our main motivation for the 
study of this galaxy, as part of our analysis of ringed non-barred
galaxies
(Verdes-Montenegro et al. 1995, 1997, 2000).
However, based on optical images, we have found that
the outer ring is just a not very well defined pseudoring lying 
 in the 
faint outer parts of the optical disk. Furthermore, 
when compared with the statistical study
performed by Athanassoula et al.
(1982) we find that
the size ratio of the outer and inner ring is far larger
than the one needed to link them to the ratio of a ring
at the outer Lindblad radius and a ring near corotation
due to a single pattern speed.

New HI data of this galaxy
show that its outer disk is warped, and that 
star formation occurs beyond  R$_{25}$ (at 1.4 $\times$ 
R$_{25}$,
where R$_{25}$ = 161\arcs\ $\pm$ 3\arcs\ 
from RC3). Star formation in the faint outer  parts of galactic disks
is a relevant and little understood issue: 
 very few examples  have been so far studied 
(Ferguson et al. 1998 and references therein).
These combined characteristics make NGC 3642 an interesting galaxy, 
and in this paper we present a detailed optical and HI 
(WSRT) study.

The systemic heliocentric radial velocity of this galaxy, obtained
from our HI data (Sect. 2.3) is 1572 km s$^{-1}$,
and corrected to the centroid of the Local Group is
1682 km s$^{-1}$. A Hubble constant of 75 km s$^{-1}$ Mpc$^{-1}$
gives a       
distance of 22.4 Mpc for NGC 3642.

\section{Observations and data analysis}
 
\subsection{Photometry}

We have analysed optical data of NGC 3642 in the B and R bands.
Boroson (1981) has obtained photographic surface photometry of this
 galaxy in the B-band.
We have supplemented Boroson's data with CCD surface photometry
extracted from an R-band image obtained from the ING archive.
This image was obtained on February 3, 1995 at the INT telescope 
with a 1024 $\times$ 1024 pixel
TEK1 CCD camera, each pixel being 24$\mu$m in size and 0.58\arcs .
The measured seeing  is 1\arcsper 7.
Flatfielding was achieved by producing
a superflatfield from the median of all obtained sky flats.
Several fields of Landolt phometric standards were also retrieved
from the ING archive, for the calibration of the R image.
We find good agreement with the profile
given by Boroson for the inner parts. For the 
outer parts some disagreement is found and we checked that this is
not
due to a bad sky subtraction of our R image.

We have also inspected the HST image published by Barth et al. (1998) 
with the F547M (V band) filter, which we retrieved from the
archive.
This image does not extend with sufficient 
 signal to noise ratio to the inner ring
but is well suited to explore the central part of NGC 3642, including
the
nuclear structure.    
 
\begin{figure}
\includegraphics[width=8.5cm]{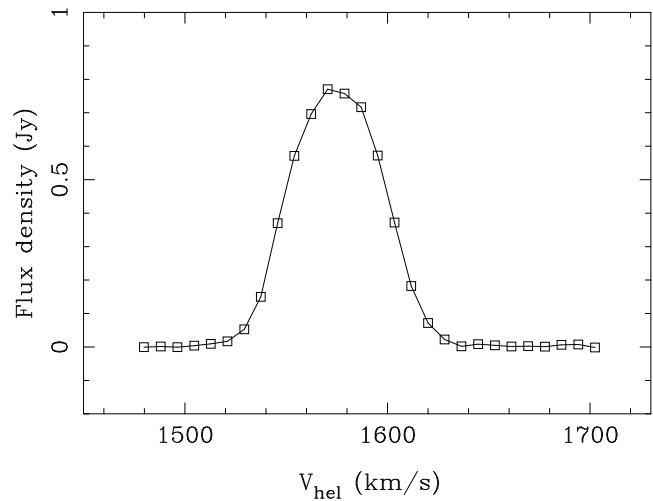}
\caption{HI flux density of NGC 3642 as a function of the heliocentric
velocity.}
\end{figure}

\begin{figure*}
\caption{(a) R band image of NGC 3642 in a grey scale 
selected to show the outer spiral structure. The dashed
 box indicates the area
shown in panel (b).
The ellipse corresponds to the outer ring size given by DVB80, 
$\sim$ 5\arcm\ $\times$ 4\arcm.
(b) The same R image as in (a) where
the central area of the galaxy is shown. Isophotes at
19.6, 19.4, 19.1, 18.8, 18,6, 18.1 and 17.8 mag/arcsec$^2$ are
plotted 
also. The dashed
 box indicates the area
shown in panel (d).
(c)   Sharpening of the R image obtained  by  the
subtraction of a 9\arcs\, $\times$ 9\arcs\,  box median filtered
image.  Darker areas
correspond to excess emission.
(d) Close-up of the sharpened image shown in (c) where the inner ring
can be seen. The major and minor axe of the inner ring are marked.
In the lower left inset we show the central part of the HST-V  image
(Sect. 2.1) , sharpened
in order to show the single arm structure. It has been obtained  by  the
subtraction of a 7\arcs\, $\times$ 7\arcs\,  box median filtered
image to the original one.  Darker areas
correspond to excess emission.
The orientation of all the images is North up and East to the left,
and the x
and y axe are plotted as offset with respect to the optical 
center of the galaxy (Table 1).}
\end{figure*}

\subsection{Radio observations}

We have observed NGC 3642 in the 21 cm line of HI with 
the Westerbork Synthesis Radio Telescope (WSRT\footnote{The  
Westerbork Synthesis Radio Telescope is operated by the Netherlands 
Foundation for Radio Astronomy with financial support from the 
Netherlands Foundation for the Advancement of Pure Research 
(N. W. O.)}) in 1983. 
We used 40 interferometers with spacings ranging from 36 m to 2844 m
in steps of 72 m. 
This results in a synthesized beam of 13\arcsper 0  $\times$
15\arcsper 1
($\alpha \times \delta$) and first grating response at 10$^{\prime}$ 
$\times$ 12$^{\prime}$ ($\alpha \times \delta$). 
We used a digital backend (Bos et al. 1981) resulting in 31
channel maps at heliocentric 
velocities 1463.3 to 1710.7 km s$^{-1}$. Hanning smoothing was 
applied on--line, giving a velocity resolution of twice the
channel spacing of 8.2 km s$^{-1}$.

The data were edited and calibrated as explained in Verdes-Montenegro
et al. (1995). 
A rms noise level of \approx \ 1.13 mJy/beam
was achieved after 12 hours of
integration. In order to get a higher signal--to--noise ratio in the 
integrated HI
distribution and associated radial velocity field, 
we
convolved the map data with a 
gaussian, leading to a beam size of 
21\arcsper 4 $\times$ 18\arcsper 4 ($\alpha \times \delta$).
Primary beam corrections have 
been applied to our maps.

The total HI emission in this galaxy was measured to be
66.16 $\pm$ 5.98 Jy \kms\ by Haynes \& Giovanelli (1991) with 
the Green Bank radiotelescope
and corrected for source to beam
extent to 83.84 Jy \kms  . The HI spectrum 
obtained by integrating the flux density in each channel map over an 
area containing the line emission is shown in
Fig. 1, and the derived total flux is  44 Jy \kms .
 Hence we are missing emission, and comparison with the single dish 
spectrum indicates that
it comes from those velocities where the emission is more extended
 (velocity range 1545 - 1590 \kms , see Sect. 3.2). 
From our spectrum we derive a systemic velocity, obtained 
as the mean velocity at 20\% of the peak, of 1572\kms . The spectrum
has 
a FWHM at the 20\% level of 74\kms .

\section{Results and discussion}

\subsection{Optical emission}

In Fig. 2a we show the R band image of NGC 3642. The 
central parts have been saturated
in order to show better the outer spiral structure, that extends up
to 
220\arcs . We have marked with an ellipse
the outer ring size given by DVB80, 
$\sim$ 5\arcm\ $\times$ 4\arcm. We do not find a complete
outer ring in our image, 
although part of the ellipse seems to coincide with the wide outer 
spiral structure, that extends from about 145$^{\circ}$ to
325$^{\circ}$.
 It is not obvious whether the thinner structure that extends from 
about 90$^{\circ}$ to 145$^{\circ}$ at smaller radius is part of 
the wider structure,
or whether it corresponds to an inner spiral. 

Boroson (1981) described the galaxy as  having ``at first     
glance a large bulge and a very faint disk'', but indicated that 
 ``after closer examination it turns that much of the bulge region 
shows spiral structure,
and is actually a section of the disk with a shorter scalelength''.
This (flocculent) spiral structure is 
shown in Fig. 2b, extending between
radius 20\arcs\ and 50\arcs , where the faint outer disk starts.

\begin{figure*}
\caption{R band radial brightness profile calculated by integrating
the deprojected image over annuli circular in the plane of the galaxy,
and with thickness of 1\arcsper5, after star subtraction.  The upper
right inlay shows the location on the image of the areas corresponding
to the bumps in the profile at radius 110\arcs\ and 170\arcs .}
\end{figure*}

The  spiral structure of NGC 3642 has a larger contrast in the outer
parts 
than in the inner parts, and the difference 
between the two parts  is more evident after 
substraction of a smoothed R image from the original one,  that
enhances
the small scale structure (Fig. 2c).
DVB80 reported the existence of an inner ring 
with a size of 25\arcsper 8 $\times$ 22\arcsper 8. This ring is faint
in our
R band image, and enhancement with unsharp masks is required in
order to isolate it (Fig. 2d).
  We have measured its size as 28\arcsper 6
 $\times$ 
25\arcsper 4,  in good agreement with DVB80, 
and its position angle  as 0$^{\circ}$.

The third ring reported by  DVB80, a nuclear one  with a diameter of
6\arcs ,
cannot be resolved in our R band image, where the central part is 
saturated due to the bright LINER nucleus (Heckman 1980). 
The inner 20\arcs\ of NGC 3642 have bright emission in the HST-V
image, 
that shows sections  (three quarters of a circle) 
of what could be the nuclear ring, with a diameter of 5\arcs . 
However a sharpening 
of the HST-V
image (Fig. 2d)  suggests  that instead of a ring this emission
enhancement is part of the 
nuclear spiral structure, since  a single m=1 spiral is observed,
reaching the active nucleus,
and extending between 1\arcs\ and 6\arcs\ from the center (109 to 654 pc at
22.4 Mpc), with several dust lanes crossing the outer part of the
minispiral.  At larger radii  the arm seems to enter the
 flocculent structure that we have indicated above.

Hence  NGC 3642 exhibits three different types of spiral
structure:
a nuclear one-armed minispiral, an inner flocculent spiral and an
outer 
 spiral that winds into an outer pseudoring extending for about half 
a circle. The winding direction of all three is the same
(clockwise).

The radial   brightness profile shown in Fig. 3 
 has been   calculated from the R band by integrating over circular
annuli with
          thickness of 1\arcsper5 on the deprojected image of the
galaxy
(see Table 1 for the deprojection parameters), 
after blotting of the stars that project on the galaxy.
Two bumps are evident on the profile at radii of
 110\arcs\ and 170\arcs , respectively, both 
associated with bright features in the spiral
structure, as indicated in the upper right inlay.
It is not surprising that such features appear in the R band since 
it includes 
the H$\alpha$ line.

\begin{figure*}
\centering
\includegraphics{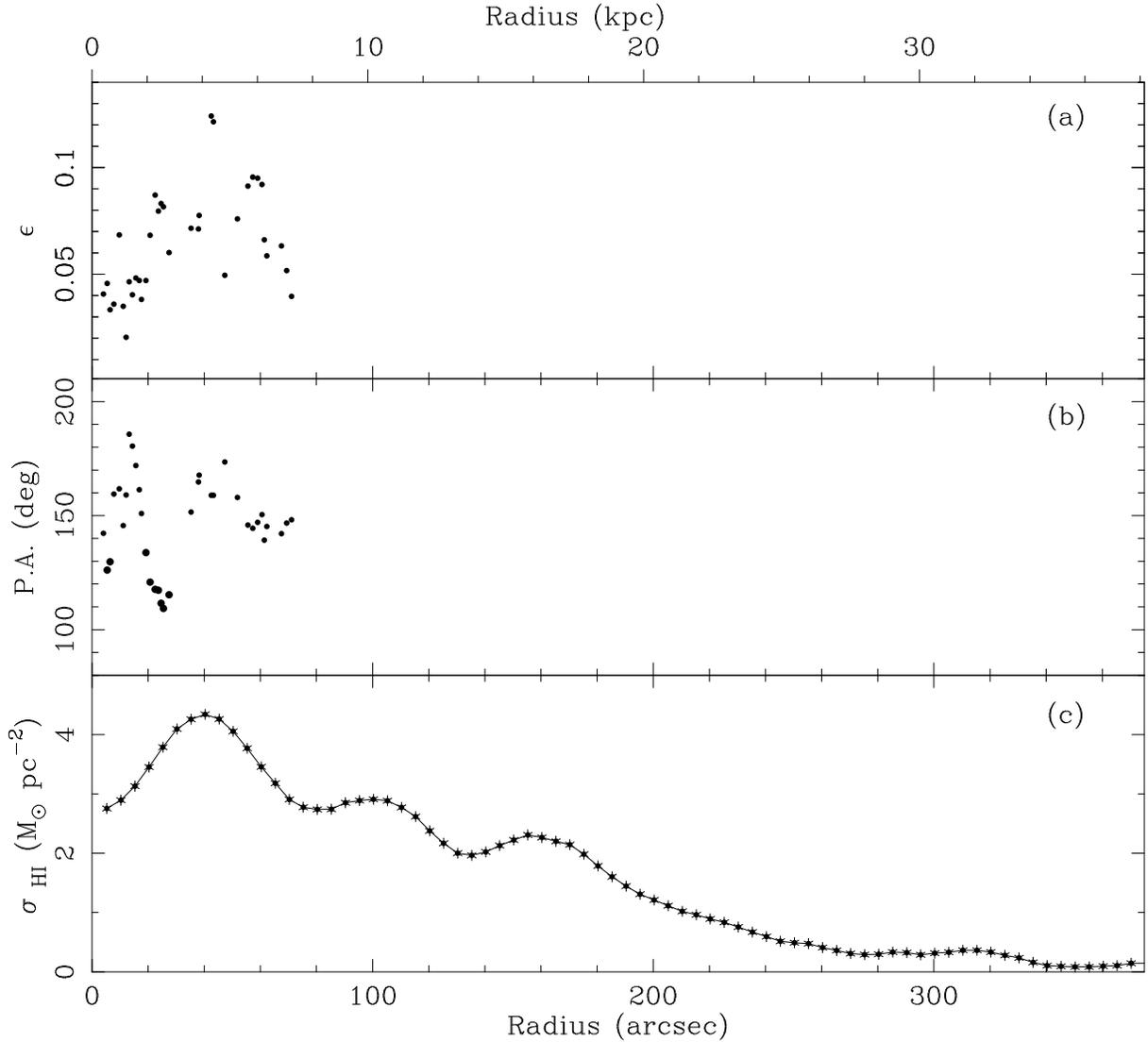}
\caption{Ellipticity (a) and position angle (b) of the ellipses fitted
to the R band isophotes as a function of their semimajor axis
length/radius (see Sect.  3.1). Angles are measured from N to E. (c)
HI surface density radial profile obtained as explained in sec. 3.2.
The lower radial scale is in arcsec and the upper one in kpc, for D =
22.4 Mpc.}
\end{figure*}

Ellipticities ($\epsilon$) and position angles (p.a.) of the 
ellipses fitted to the  R band isophotes are shown in 
Fig. 4a and b respectively as a function of radius, 
up to the larger radius at which  
we were able to perform such fitting (radius of 70\arcs ).
The results are very noisy since the  isophotes
are strongly influenced by the spiral arms. 
The ellipticity 
shows a mean value of 0.06 $\pm$ 0.02, corresponding to 
an inclination of $i$ = 20.4$^{\circ}$ $\pm$ 
3.5$^{\circ}$, and the position angle does not show
particularly abrupt changes with radius as expected for
a bar embedded in a disk.
There is also no obvious difference of the value of the
inclination as function of radius, although the inner parts
seem a bit more circular. Furthermore, there are no clearly
marked minima in the light distribution corresponding to the 
Lagrangian points L$_4$ and L$_5$, a characteristic of the
morphology of an oval distortion, nor is there any other
sign of the presence of a bar.
The low inclination makes the morphological determination of the
position angle highly uncertain. Inside the inner pseudoring
 (r $<$ 14\arcs ) the p.a. ranges between 
125$^{\circ}$ and 160$^{\circ}$. 
From 14\arcs\ to 20\arcs, i.e. up to the 
beginning of the spiral structure, the p.a. drops steeply
from 180$^{\circ}$ down to 110$^{\circ}$.
Then it jumps up to $\sim$ 158$^{\circ}$ $\pm$ 44$^{\circ}$, between 
20\arcs and 70\arcs .

\begin{table}
\caption[ ]{Parameters of NGC 3642.}
\begin{tabular}{lc}
\hline 
Center position$^a$&  \\
$\alpha$(1950.0)&  11$^h$ 19$^m$ 25\arcsgper 0  \\
$\delta$(1950.0)& 59$^{\circ}$ 20$^{\prime}$ 54\arcsper 1\\
Inclination ($^{\circ}$) ~~~~~ adopted& 20.4$^{b}$ $\pm$ 3.5, \\
~~~~~~~~~~~~~~~~~~~~~~~~~~ others &  2$^{d}$, 
34$^{e}$, 
25$^{f}$,19$^{g}$ \\
Position angle ($^{\circ}$)~~ adopted& 122.6$^{c}$  \\
~~~~~~~~~~~~~~~~~~~~~~~~~~ others& 105$^{d,e}$, 
100$^{f}$  \\
B$_T$$^e$& 11.65 \\
B$_T^o$$^e$&  11.61 \\
 R$_{25}$$^e$& 161\arcs\, (17.4 kpc)\\
HI systemic heliocentric velocity (km s$^{-1}$) & 
1572$^c$,1588$^e$
1586$^h$\\
\hline
\end{tabular}
\begin{list}{}{}
\item[$^{\rm a}$]  Central position of NGC 3642 from Cotton et al.
(1999).
\item[$^{\rm b}$] This paper, from ellipse fitting to the R band
image.
\item[$^{\rm c}$] This paper from HI velocity field.
\item[$^{\rm d}$] UGC catalog.
\item[$^{\rm e}$] RC3.
\item[$^{\rm f}$]  Boroson (1981).
\item[$^{\rm g}$] Kornreich et al. (1998).
\item[$^{\rm h}$] Haynes \& Giovanelli (1991).
\end{list}
   \end{table}

\begin{figure*}
\centering
\caption{Channel maps of the 21 cm line radiation superimposed on the R image,
with the central part blotted for clarity. The
heliocentric velocities are
indicated in each panel. Contours correspond to 
4.6, 7.6, 10.7, 13.7, 16.7 and 19.7 K, and the
rms noise of the maps is 1.9 K. The synthesized beam (21\arcsper 4 
$\times$
18\arcsper 4  -- $\alpha \times \delta$) is plotted in the upper 
left panel.}
\end{figure*}

\begin{figure*}
\centering
\caption{(a) Contour map of the HI column density distribution in NGC 
3642 overlapped on the R image. 
The contours are 1.1, 3.4, 5.6, 7.9, 10.2, 12.4, 14.7, 16.9, 19.2 and
21.4 
 $\times$ 10$^{20}$ 
atoms cm$^{-2}$.
(b) Greyscale map of the  HI column density distribution with
contours
as in (a). The main features are marked as dots.
The synthesized beam (21\arcsper 4 
$\times$
18\arcsper 4  -- $\alpha \times \delta$) is plotted in the upper left
of both
panels.}
\end{figure*}

\begin{figure*}
\centering
\caption{(a)  Map of the first--order moment of the radial velocity 
field where both iso-velocity contours and greyscale are shown for
clarity.
The scale goes as in the wedge, where the numbers indicate
heliocentric
velocities in km s$^{-1}$. The contours go from 1532 to 1602 \kms\
with a
step of 5 \kms .
 The main spiral features are marked as dots. The straight line
indicates
 the direction of the position-velocity cut shown in Fig. 8.
The beam size is 21\arcsper 4 $\times$ 18\arcsper 4 and is plotted in
the upper left.
      (b) Central part of the velocity field shown in (a). The major
and minor
axis directions are indicated, and a cross indicates the optical
center 
position (see Sect. 3.2).}
\end{figure*}

\subsection{Atomic gas distribution and kinematics}

In Fig. 5 we display the channel maps containing the HI emission at
the indicated heliocentric velocities, produced with a beam of 
21\arcsper 4\ $\times$ 18\arcsper 4. Each channel has a width 
of 8.2\kms\ and an  
rms noise of 1.9 K, implying a level 
of 3.4 $\times$ 10$^{19}$ cm$^{-2}$. 
The HI contours have been superimposed on the R-band image. 
There is a good correspondence between the brighter areas of
HI emission and the  outer spiral structure, as is also seen in 
the integrated HI map  (Fig. 6).
 The brighest 
HI/optical features have column densities of 2.7 $\times$ 10$^{21}$ 
cm$^{-2}$. At a level of 1.5  $\times$ 10$^{20}$ cm$^{-2}$
the galaxy extends $\sim$ 240\arcs\ in radius (1.5  $\times$
R$_{25}$), 
offcentered by $\sim$ 25\arcs\ to the west.
This is plausibly a lower limit to the real extent since we are
missing
nearly half  of the flux (see Sect. 2.3).
 There is a central HI depression with N(HI) $\sim$ 1.0 $\times$ 
10$^{19}$ cm$^{-2}$.
 The azimuthally
       averaged radial distribution of the HI column density is shown
in 
Fig. 4c. It has been obtained by averaging the
 two-dimensional HI distribution in elliptical rings using the
geometrical
 parameters given in Table 1. 

Numerous channel maps show emission at both receding and approaching 
sides of the galaxy. 
This results in the velocity field shown in
 Fig. 7a, where redshifted emission is observed in the blueshifted
part
of the velocity field and viceversa.
Between 40\arcs\ and 70\arcs\ the isovelocity contours 
show kinks that correlate with the HI clumps associated to the spiral
arms. 
But most of the perturbations for the largest radii are
characteristic
of a warped disk.
 We have 
derived the kinematical major axis based on the most symmetrical part
of the velocity field (r $<$ 40\arcs ). The major axis has been 
obtained 
as the direction perpendicular to the isovelocity contours, resulting
in 
a value of p.a. = 122.6$^{\circ}$. This direction coincides well with
the line joining the peak of the 
 blueshifted and redshifted velocities in the 
inner parts (Fig. 7). It differs significantly from the optically 
derived one, although this is not surprising due to the low
inclination
and the effect of the spiral structure on the isophotal fitting. We
will thus
adopt the p.a. of NGC 3642 derived from the velocity field.
 The isocontour at the systemic velocity 
 (1572\kms , Sect. 2.3) intersects  the major 
axis direction at the optical center position, which we will
also adopt as the kinematical center of the galaxy (Fig. 7b).

A cut along the major axis direction (Fig. 8, filled circles) 
shows clearly the radial 
velocity inversion observed in the velocity field.
As we show in the next section, this can be modelled as a warped
disk:
due to the low inclination of the galaxy, any small difference in the 
inclination between the inner and outer parts can produce a large 
rotation of the major axis.
We have overlapped the velocities obtained from a cut along the major
axis
of the velocity field. Declining velocities are 
observed  for radii larger than 100\arcs\ 
down to values even lower than the systemic velocity.

\subsection{Description of the warp model}

 We have modelled the velocity field 
                fitting a tilted ring model (see Begeman 1987; the
ROTCUR
and GALMOD 
          tasks in Gipsy). 
 We have tried different combinations of the
          inclination, position angle and rotational velocity, as we
          describe below.
          In all cases the galaxy has been divided into concentric
          rings, each of them with a width of 5\arcs\ along the major
axis
and a central position fixed to the optical center  (Table
          1).
Points within a sector of 30$^{\circ}$ from the minor axis were
          excluded from the fits. 
 The
          expansion velocities were set to zero and the systemic
velocity
          fixed to the central velocity of the HI spectrum
         (Sect. 2.3, Table 1).

We could not allow the
inclination and
          position angle to vary freely, 
since the errors for the
          inclination and for the
rotation velocity are large for all
          radii. 
The galaxy is seen so face-on that a determination of the rotation
curve is almost impossible : the uncertainties of the inclination
correction are simply too large. Nevertheless, we have attempted
to model the galaxy, assuming a flat rotation curve, as a
continuation
of what we find in the inner parts. From this we can estimate an
approximate
amplitude of the warping.
In Fig. 9 we show the best combination of position angle and
inclination
that we have found in order to reproduce the velocity field. The
inversion
in the velocity field is reproduced although we could not obtain its
most extreme
features (Fig. 10). 
The fit along the major axis is however very accurate as shown in
Fig. 8, where 
we compare  the rotation curve along the major axis for the model and
original data.
The rotation curve, inclination and position angle obtained from our 
modelling are plotted in Fig. 9. 
The inclination derived from the modelling of the HI emission fits
well
with the one derived on the base of R band isophotal fitting.
A warping of the disk of $\sim$ 25$^{\circ}$
is needed in order to reproduce the velocity field of NGC 3642.
The bumps in the position angle at $\sim$ 90\arcs\ and 140\arcs\
coincide
with the location of the spiral arms.

An estimate of 
the rotational velocity of this galaxy can be obtained from 
the Tully-Fisher relation
(Pierce \& Tully 1992). Using the observed $M_B$ we get
$W_R$ = 379 \kms , which, combined with the observed extrema of
the radial velocities, implies an inclination of 17$^{\circ}$,
in good agreement with our adopted value given the scatter in the 
Tully-Fisher relation.

\subsection{Comment on the z-velocities in warps}

The galaxies most useful for 
 rotation curves studies are inclined typically 
between 50\degr\ and 80\degr\ (e.g. Begeman
et al. 1991), and if a warp is present in the outer parts, its effect
 is traditionally modelled using a
tilted ring model (Rogstad et al. 1974, Bosma 1978, 1981, and Begeman
1987). Such a procedure, in particular the one
implemented in ROTCUR, minimizes the importance of any peculiar
motions, and maximizes the circular velocity in each ring independent
of
its spatial orientation. In reality, there ought to be some
component of the motion of the inclined outer rings towards the
main plane defined by the inner parts of the galaxy. We can get an
upper limit to this motion by attributing to it all the peculiar motion
seen in Fig. 8 (p-v-diagram along the major axis), which is
30  \kms . This can be compared to the rotation velocity, 
which is certainly larger than  80\kms , and more likely of the order of 150
\kms\
 or even 190 \kms\  (if the TF relation holds and the Sb(r)I
classification 
of the RSA is retained). 
So at first order, z-motions of warps can be neglected.

\begin{table}
      \caption{Derived Parameters. }
\begin{tabular}{lc} 
\hline
H$_I$ flux$^a$& 44 Jy km s$^{-1}$\\
H$_I$ flux$^b$& 84  Jy km s$^{-1}$\\
M$_{HI}^a$& 5.2 $\times$ 10$^9$ M$_{\odot}$\\
M$_{HI}^b$& 9.9 $\times$ 10$^9$ M$_{\odot}$\\
M$_{HI}$/L$_B^0$& 0.56\\
\hline
\end{tabular}
\begin{list}{}{}
\item[$^{\rm a}$]  
This paper
\item[$^{\rm b}$]  Haynes \& Giovanelli (1991).
\end{list}
   \end{table}

\subsection{Environment}

We have studied the neighborhood of NGC 3642 within a radius of 500 
kpc and 500\kms\ and show the results in Fig. 11. From
the NED\footnote{The NASA/IPAC
extragalactic database (NED) is operated by
   the Jet Propulsion Laboratory, California Institute of
Technology,
 under contract with the National Aeronautics and Space
Administration}
 database we found that NGC 3642 has  10 neighbors 
with known redshift in the 
box under consideration.  
Among them, NGC 3610 and NGC 3613 have magnitudes similar to
that of 
NGC 3642, while the rest are fainter by more than 2 magnitudes.
  The closest companion,
NGC 3610,  is located 
at an apparent separation of 34\arcmper 7, or 226 kpc, and 
has clear signs of merger:
this is a quite perturbed elliptical galaxy, 
with  a prominent, edge-on stellar disk along the major
       axis in the inner regions (Scorza \& Bender 1990). It 
has several shells, plumes and 
boxy outer isophotes indicating a past merger involving two disk
galaxies
dated 4 Gyr ago (cf Whitmore et al. 1997) based on UBV colors.
 NGC 3613 is a disky elliptical with the disk fully embedded in a 
spheroidal component (Michard \& Marchal 1994).

\begin{figure}[h]
\centering
\includegraphics[width=8.5cm]{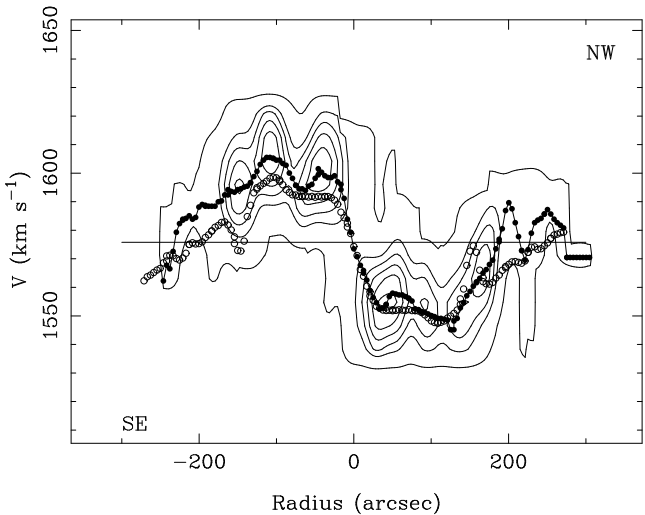}
\caption{(a) Position-velocity diagram obtained across the major axis
direction
of the HI data cube. 
The plotted levels are 1.5, 6.1, 10.7, 15.2, 19.8, 24.4, 28.9, 33.5
and 
38.1 K. Filled circles correspond to a cut  along the observed
velocity field (Fig. 7a), while the empty ones correspond to a cut 
along the modelled  velocity field.
The synthesized beam is 21\arcsper 4 
$\times$ 18\arcsper 4  -- $\alpha \times \delta$.}
\end{figure}

\begin{figure*}
\centering
\includegraphics{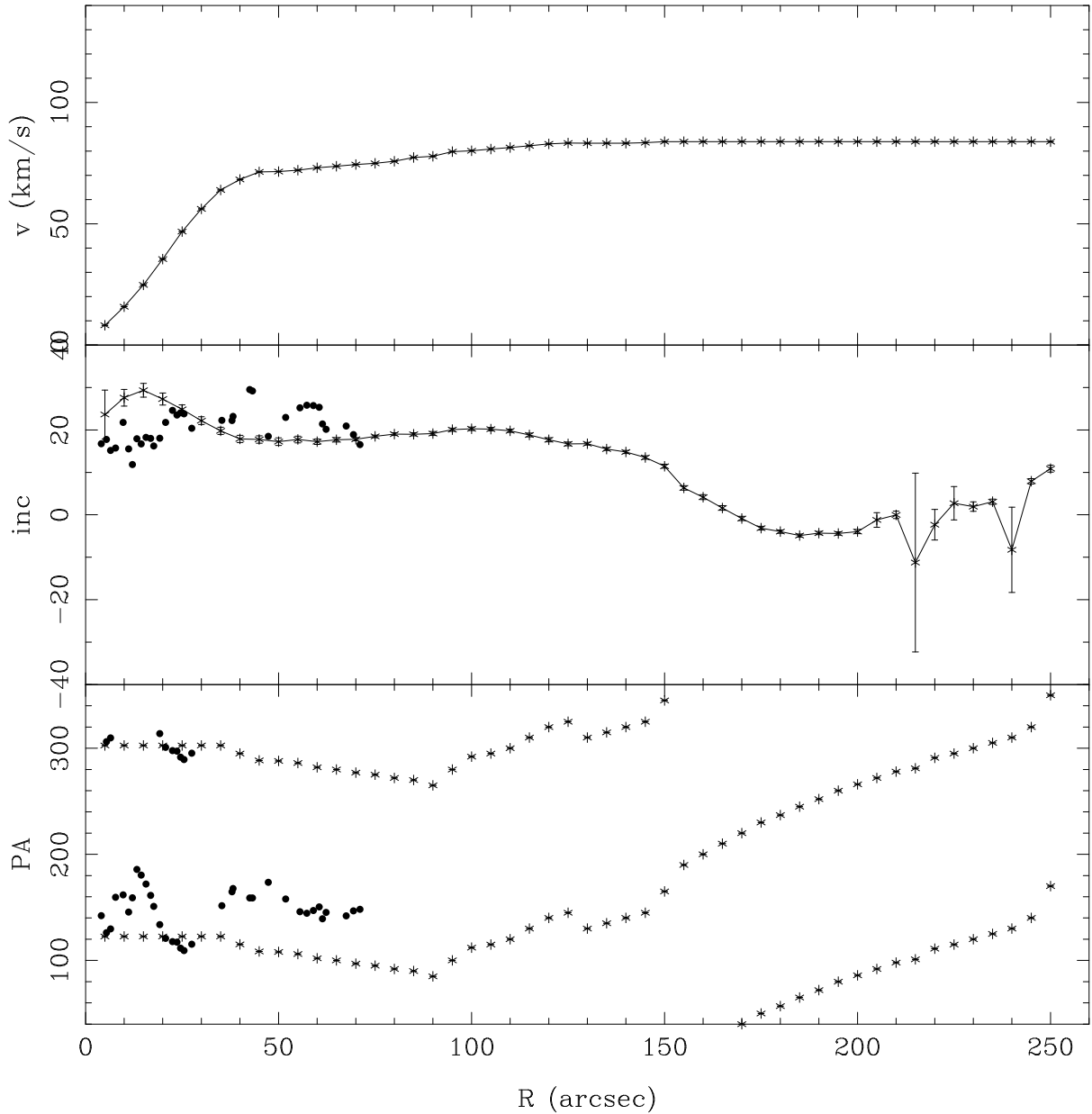}
\caption{(a) Rotation curve assumed for the modelling of the
velocity \
field. The best fit to this curve and to the observed velocity field
is given by the combination of inclination and position angle
plotted in (b) and (c) respectively.
Filled dots correspond to the isophotal fitting of the optical R
image (Fig. 4). 
Angles are measured from North to East. }
\end{figure*}

 Garcia (1993) considered a larger scale, and finds that 
NGC 3642 belongs to a
group      
composed by 5 galaxies. The brightest members of this group are 
NGC 3642 and 3610, followed by
NGC 3619, a magnitude fainter, and by NGC 3674 and NGC 3683,
which are two magnitudes
fainter. The second nearby companion of NGC 3642 in
mass, NGC 3619, is an early type disk galaxy with shells (cf.
Schweizer
\& Seitzer 1988, Van Driel et al. 1989).
The velocity dispersion of the group is 600 \kms . NGC 3683a
is 
close to the group, although with a larger redshift (see Fig. 11).

\section{Discussion and conclusions}

We have investigated the optical morphology as well as the
 gas distribution and kinematics of the nearly
face-on spiral galaxy NGC 3642.  As is typical for  spiral galaxies,
the 
HI gas extends a factor of 1.5 beyond the optical radius.
 What is more unusual
is that  the outer HI regions composing the  pseudoring are warped. 
Furthermore the HI is more extended in the western side, where the more
external
star forming area is located, and higher column densities are
associated
to these regions. These perturbations affect only to the 
outer disk, since 
 the
kinematics within the main body conforms well to an ordinary 
differentially rotating disk.

The inner kpc of NGC 3642 shows a one-armed spiral structure. 
 Spirals in the inner kpc of active galaxies have been
considered as drivers of accretion into the central engine (Regan \& Mulchaey 
1999). While
barred galaxies tend to show bisymmetrical patterns (Athanassoula
1992) non barred ones can have more irregular structures. The observed
spiral could be driving nuclear accretion.

Both phenomena, nuclear activity and star forming warped outer parts,
might be linked to the fact that  NGC 3642 is located in a rich
environment.
Given the clear signs of merging in its closest bright 
companions, 
 it is reasonable to assume 
that NGC 3642 also underwent recently an interaction, or, more likely,
captured a low-mass, gas-rich dwarf, which disintegrated, and whose
debris now form the outer HI and the outer patchy spiral structure
in NGC 3642 itself. The differences in the kinematics 
between the approaching and
receding sides
of the galaxy
are probably due to the accretion of a companion.
 The low surface brightness extension of the galaxy,
as seen in the photometric structure, also points in this direction.
Such low surface brightness extensions have been studied before, (e.g.
Longmore et al. 1979, Bosma \& Freeman 1993), and are not so
uncommon.
Longmore et al. (1979) indicated already that under favorable initial
conditions,
specially a low relative velocity encounter, a considerable fraction
of the gas content of a galaxy can be drawn out in an interaction, 
and yet participate in subsequent star formation. 
We could therefore see star formation
 in the denser regions of redistributed material. i.e. in the outer
pseudoring, which  might be in the process of forming a continuous
ring. 

In conclusion, the observed outer and perhaps even the nuclear 
structure of the unbarred galaxy NGC 3642 can be
attributed 
to  accretion of an HI-rich dwarf galaxy that could be acquired
from its rich environment. Star formation was triggered in its
outskirts
by incorporation of external gas that produced also a pronounced warp
in the gaseous disk. Our modeling of the velocity field of this
face-on
galaxy allowed us to set a constrain on the amplitude of 
the  z-motions in warps, 
which we find to be smaller than about
30\kms . It could well be that the formation of some of the
faint outer extensions
of galactic disks
is due to this kind of destruction of a low-mass, gas-rich dwarf
during capture.

\begin{figure}[t]
\centering
\caption{First-order moment of the
 modeled channel maps 
obtained with the geometrical parameters plotted in Fig. 9.}
\end{figure}

\begin{figure*}[h]
\centering
\includegraphics{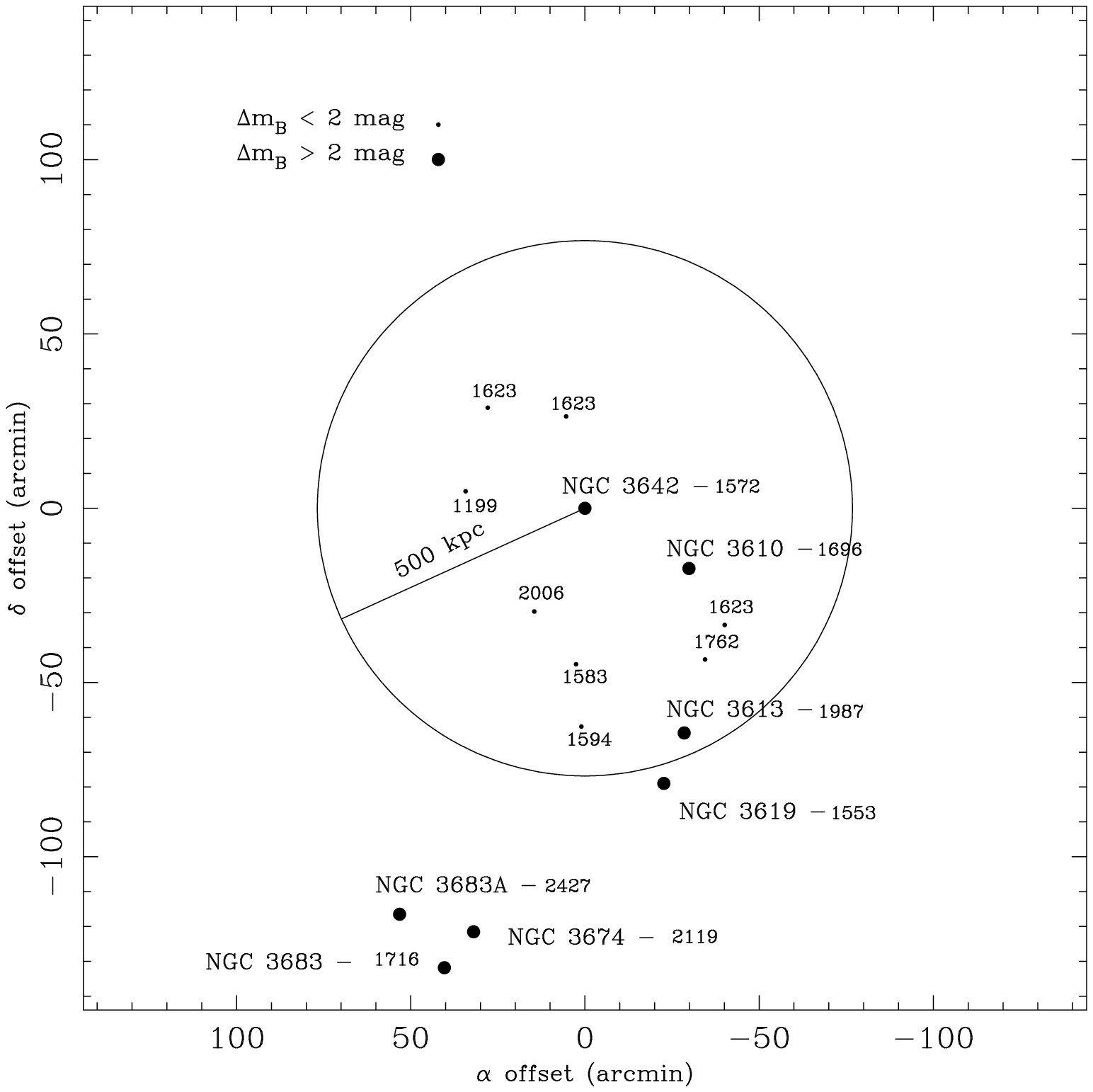}
\caption{The enviroment of NGC 3642. The circle encloses an area of 
500 kpc radius centered on the galaxy. Those galaxies with 
a difference in blue magnitude less than 2 mag with respect to NGC
3642 
are marked with larger circles, while the rest are marked with
smaller ones.
The velocities are also indicated.}
\end{figure*}

\newpage

\begin{acknowledgements}
LV--M acknowledges the hospitality and financial support 
of the Observatoire de Marseille,  
where part of this work was carried out. 
LV--M is partially supported by AYA 2000-1564 
 and Junta de Andaluc\'{\i}a (Spain).
\end{acknowledgements}

\end{document}